\documentclass[12pt]{article}
\usepackage{epsfig}
\usepackage{amsfonts}
\usepackage{amscd}
\usepackage{latexsym}
\usepackage{amsmath,amssymb}
\usepackage{verbatim}
\usepackage{setspace}
\usepackage{color}
\usepackage{cite}

\usepackage[textheight=9in, textwidth=6.5in, letterpaper]{geometry}

\usepackage{hyperref}
\hypersetup{
    colorlinks,
    citecolor=black,
    filecolor=black,
    linkcolor=black,
    urlcolor=black
    linktoc=all
}

\numberwithin{equation}{section}

\def\e{{\epsilon}}

 \def\p{\partial}
 \def\bz{{\bar z}}
 
\def\0{{(0)}}
\def\1{{(1)}}
\def\2{{(2)}}

\def\ci{{\mathcal I}}

\def\<{\langle }
\def\>{\rangle }

\newcommand{\bea}{\begin{eqnarray}}
\newcommand{\eea}{\end{eqnarray}}
\newcommand{\be}{\begin{equation}}
\newcommand{\ee}{\end{equation}}
\newcommand{\ba}{\begin{align}}
\newcommand{\ea}{\end{align}}
\def\be{\begin{equation}}
\def\ee{\end{equation}}
\def\beq{\be\begin{array}{c}}
\def\eeq{\end{array}\ee}
\def\Phi_1{E_r }

\renewcommand{\epsilon}{\varepsilon}

   \makeatletter
  \let\over=\@@over \let\overwithdelims=\@@overwithdelims
  \let\atop=\@@atop \let\atopwithdelims=\@@atopwithdelims
  \let\above=\@@above \let\abovewithdelims=\@@abovewithdelims
\renewcommand\section{\@startsection {section}{1}{\z@}%
                                   {-3.5ex \@plus -1ex \@minus -.2ex}
                                   {2.3ex \@plus.2ex}%
                                   {\normalfont\large\bfseries}}

\renewcommand\subsection{\@startsection{subsection}{2}{\z@}%
                                     {-3.25ex\@plus -1ex \@minus -.2ex}%
                                     {1.5ex \@plus .2ex}%
                                     {\normalfont\bfseries}}

\begin{document}
\begin{titlepage}
\unitlength = 1mm

\ \\
\vskip 1cm
\begin{center}

{ \LARGE {\textsc{New Symmetries of QED}}}

\vspace{0.8cm}
Daniel  Kapec, Monica Pate, and Andrew Strominger

\vspace{1cm}

{\it  Center for the Fundamental Laws of Nature, Harvard University,\\
Cambridge, MA 02138, USA}\\

\begin{abstract}
The soft photon theorem in $U(1)$ gauge theories with only massless charged particles has recently been shown to be the Ward identity of an infinite-dimensional asymptotic symmetry group. This symmetry group is comprised of gauge transformations which approach angle-dependent constants at null infinity. 
In this paper, we extend the analysis to  all $U(1)$ theories, including those with massive charged particles such as QED.    \end{abstract}

\vspace{1.0cm}

\end{center}

\end{titlepage}

\pagestyle{empty}
\pagestyle{plain}

\def\vx{{\vec x}}
\def\p{\partial}
\def\po{$\cal P_O$}

\pagenumbering{arabic}

\tableofcontents
\section{Introduction}

The soft photon theorem \cite{low54,low,bk,ggm,Weinberg:1965nx} has played a ubiquitous role in 
the study of QED and more general abelian gauge theories. For example, it is essential for taming otherwise uncontrollable infrared divergences in the ${\mathcal S}$-matrix and is central to the analysis of jet  substructure. Recent considerations \cite{He:2014cra, Kapec:2014zla,Mohd:2014oja,Larkoski:2014bxa,Bern:2014vva,Balachandran:2014hra,Balachandran:2013wsa,Frohlich:1979uu,Frohlich:1978bf} have demonstrated that, in abelian gauge theories with only massless charged particles, the soft theorem is a Ward identity of an infinite-dimensional symmetry group comprised of certain
`large'  gauge transformations which do not die off at infinity.  These symmetries are spontaneously broken and the soft photons are the Goldstone bosons. This is but one instance of a recently-discovered universal triangle connecting soft theorems, symmetries and memory in gauge and gravitational theories \cite{Strominger:2013lka,Strominger:2013jfa,Mohd:2014oja,Larkoski:2014bxa,Bern:2014vva,He:2014laa,Kapec:2014opa, He:2014cra, Lysov:2014csa, Kapec:2014zla, Kapec:2015vwa, He:2015zea, Strominger:2014pwa, Pasterski:2015tva, Pasterski:2015zua, Banks:2014iha,Cachazo:2014fwa,Adamo:2014yya,Geyer:2014lca,Campiglia:2014yka,Campiglia:2015yka,Lipstein:2015rxa,Adamo:2015fwa,  Broedel:2014fsa,White:2014qia}.

Of course in the real world QED  has massive, not massless, charged particles.  Hence, it is desirable to extend 
our results to the massive case. That goal is achieved in this paper. As seen below, the massive case is rather more subtle than the massless one and requires a careful analysis of timelike infinity. 

We hope that the identification given herein of the symmetry which controls the electromagnetic soft behavior of QED and more generally, the Standard Model will have practical utility for organizing and predicting a variety of physical phenomena. 

 The outline of the paper is as follows. In section 2, we establish conventions and review relevant aspects of abelian gauge theories and their asymptotic symmetries.   In section 3, we discuss the asymptotic states, derive the Ward identity of the asymptotic symmetries, and demonstrate its equivalence to the soft photon theorem.  
 
 A key ingredient of our analysis is that, in physical applications, the electromagnetic field is generically\footnote{For instance when, as in electron-positron scattering,  the dipole moment is not constant in the far past or future.} not smooth near spatial infinity $i^0$. Rather it obeys a  matching condition near $i^0$ which identifies its value at the future of past null infinity ($\ci^-_+$) with  its value at the antipodal point on the sphere at the past of future null infinity ($\ci^+_-$). 
 In the appendix, we show in detail how this follows from the standard Lienard-Wiechert formulae.

After completion of this work, we received the eprint \cite{Campiglia:2015qka} 
by Campiglia and Laddha who independently arrive at the same conclusions. They use an elegant method involving a natural
hyperbolic slicing of Minkowski space.

\section{Abelian gauge theory with massive matter  }
We consider the theory of an abelian gauge field $A_\mu$  coupled to massive matter fields $\Psi_i$ with charges $eQ_i$, where $Q_i$ is an integer, in Minkowski space. In retarded coordinates,  the Minkowski metric reads
\be
ds^2=-dt^2+ (dx^i)^2=-du^2 -2du dr + 2r^2\gamma_{z\bz}dz d\bz,  
\ee
where $u$ is retarded time and $\gamma_{z\bz}$ is the round metric on the unit radius $S^2$ with covariant derivative $D_z$. The $S^2$ coordinates $(z, \bz)$ are related to standard  Cartesian coordinates by
\be\label{adv}
r=x_ix^i, \;\;\; u=t-r, \;\;\; x^i=r\hat{x}^i(z, \bz).  
\ee 
In retarded coordinates, future null infinity $(\mathcal{I}^+)$ is  the null hypersurface $(r=\infty,u, z, \bz)$.  

Near past null infinity $(\mathcal{I}^-)$, we work in  advanced coordinates $(v,r,z, \bz)$ with line element
\be
ds^2=-dv^2+2dvdr +2r^2\gamma_{z\bz}dz d\bz .  
\ee
Advanced coordinates are given by 
\be \label{ret}
r=x_ix^i, \;\;\;\; v=t+r, \;\;\;\; x^i=-r\hat{x}^i(z, \bz), 
\ee
and $\mathcal{I}^-$ corresponds to the null hypersurface $(r= \infty,v,z, \bz)$.  
Note in particular that the angular coordinates on $\mathcal{I}^+$ are antipodally related to those on $\mathcal{I}^-$ so that a light ray  passing through the interior of Minkowski space reaches  the same value of $z, \bz$ at both $\mathcal{I}^+$ and $\mathcal{I}^-$. We denote the future (past) boundary of $\mathcal{I}^+$ by $\mathcal{I}^+_+$ ($\mathcal{I}^+_-$), and the future (past) boundary of $\mathcal{I}^-$ by $\mathcal{I}^-_{+}$ ($\mathcal{I}^-_{-}$).   

We consider theories with a $U(1)$ gauge field strength $F=dA$ subject to the Maxwell equation \be
\nabla^\mu F_{\mu \nu}=e^2J_\nu,  
\ee 
where $J_\nu$ is the matter charge current.  This is  invariant under the gauge transformations
\be \label{gaugetransform}
A_\mu(x) \to A_\mu (x) +\p_\mu \epsilon(x), \;\;\;\; \Psi_i(x)\to e^{iQ_i \epsilon(x)}\Psi_i(x),  
\ee
where $\epsilon \sim \epsilon+2\pi$ and $\Psi_i$ is a wavefunction or field. 
Gauge transformations that vanish at infinity correspond to redundant descriptions of the same physical state and can be eliminated by a choice of gauge. However,  as in the massless case \cite{He:2014cra}, we are interested in certain angle-dependent large gauge transformations which act non-trivially on physical states.

\subsection{Asymptotics  }
 We now analyze the behavior of the theory near $\mathcal{I}^+$ in retarded radial gauge
\be
A_r=0, \;\;\;\;\; A_u|_{\mathcal{I}^+}=0.  
\ee
This gauge choice leaves unfixed a class of residual large gauge transformations parameterized by an arbitrary function $\epsilon^+(z,\bz)$ on $S^2$. These gauge transformations change boundary data at $\mathcal{I}^+$ and are to be regarded as physical symmetries of the theory. 
Near $\mathcal{I}^+$, we assume the asymptotic expansion 
\be
A_u=\sum_{n=1}^{\infty}\frac{A_u^{(n)}(u,z, \bz)}{r^n}, \;\;\;\;\; A_z=\sum_{n=0}^{\infty} \frac{A_z^{(n)}(u,z, \bz)}{r^n}.  
\ee
A similar asymptotic expansion holds for fields near $\mathcal{I}^-$.  

We are interested in scattering processes for which the initial and final states consist  of non-interacting massive charges moving at constant velocities.  Hence, we require that the only contribution to the electric and magnetic fields at future/past timelike infinity ($i^\pm,t \rightarrow \pm \infty ) $ are those fields sourced by the constant velocity massive charges,
and that the magnetic fields vanish at spatial infinity:
\be
F_{z\bz}|_{\mathcal{I}^+_{-}}=0,   \;\;\;\;\;\;\;\;\;    F_{z\bz}|_{\mathcal{I}^-_{+}}=0.
\ee

In retarded coordinates, Maxwell's equations read
\be
r^{-2}\p_r(r^2F_{ru})-\p_uF_{ru}+r^{-2}(D^zF_{zu}+D^\bz F_{\bz u})  =e^2J_{u},  
\ee
\be
r^{-2}\p_r(r^2 F_{ru})+ r^{-2}(D^zF_{zr}+D^\bz F_{\bz r})=e^2 J_r ,  
\ee
\be
 \p_r( F_{rz}-F_{uz})-\p_uF_{rz}+r^{-2}D^\bz F_{\bz z}=e^2J_z.  
\ee
Massive particles with finite energy cannot reach $\mathcal{I}$, so the matter current vanishes at this surface:
\be
J_\mu|_{\mathcal{I}}=0.  
\ee
  The leading order equation for the evolution of the gauge field along $\mathcal{I}^+$ is then given by 
\be \label{evolution}
\p_uF_{ru}^{(2)}+\p_u (D^zA^{(0)}_z+D^\bz A^{(0)}_\bz)=0.  
\ee
The free data at this order includes the boundary data $F_{ru}^{(2)}|_{\mathcal{I}^{+}_{-}}$  along with the radiative mode $A_z^{(0)}(u,z,\bz)$.

In advanced coordinates, we can perform the analogous large-$r$ expansion near $\mathcal{I}^{-}$ and obtain the leading order equation\be
\p_vF_{rv}^{(2)}-\p_v (D^zA^{(0)}_z+D^\bz A^{(0)}_\bz)=0.  
\ee
  The free data at this boundary
includes the field strength boundary data $F_{rv}^{(2)}|_{\mathcal{I}^{-}_{+}}$ along with the radiative mode   $A_z^{(0)}(v,z,\bz)$. The residual large gauge symmetry is parameterized by an arbitrary function $\epsilon^-(z,\bz)$ on $S^2$. 

\subsection{Matching near spatial infinity }
The above discussion treats the asymptotic dynamics at $\mathcal{I}^+$ and $\mathcal{I}^-$ separately. However, to study the semiclassical scattering problem, we must first specify how to relate free data and symmetry transformations at $\mathcal{I}^+$ to their counterparts at $\mathcal{I}^-$.   Generic solutions to the sourced Maxwell equations satisfy\footnote{See appendix for an expanded discussion of this matching condition.}

\be \label{matchF}
F^{(2)}_{ru}(z,\bz)|_{\mathcal{I}^+_-}=F^{(2)}_{rv}(z,\bz)|_{\mathcal{I}^-_+}.
\ee
Recalling that, according to (\ref{adv}) and (\ref{ret}), the points labelled by the same $(z,\bz)$ in retarded and  advanced  coordinates  are antipodally related,  this equates the boundary values of past and future fields at antipodal points near spatial infinity $i^0$. 
As discussed in \cite{He:2014cra, He:2015zea}, a CPT  and Lorentz-invariant  matching condition for the gauge field is given by 
  \begin{align} \label{matchA}
 	A_z(z,\bz) \big|_{\mathcal{I}^+_-} = A_z(z,\bz) \big|_{\mathcal{I}^-_+} .
 \end{align}
Requiring that  the large gauge transformations preserve this matching condition gives:
 \begin{align} \label{matchEp}
 	\epsilon^+(z,\bz) = \epsilon^-(z,\bz).  
 \end{align}
 This matching condition singles out a canonical diagonal subgroup of the large gauge transformations at $\mathcal{I}^+$ and $\mathcal{I}^-$. The corresponding gauge parameters are constant along the null generators of $\mathcal{I}$ and generate nontrivial physical symmetries of the $\mathcal{S}$-matrix.

\subsection{Mode Expansions  }
The standard mode expansion for the gauge field in the plane wave basis takes the form
\be
A_\mu(u,r,z,\bar{z})=e\sum_{\alpha}\int \frac{d^3 q}{(2\pi)^3}\frac{1}{2\omega_q}[\epsilon^{*\alpha}_\mu(\vec{q})a_\alpha(\vec{q})e^{iq\cdot x}+\epsilon_\mu^\alpha(\vec{q})a_\alpha(\vec{q})^\dagger e^{-iq\cdot x}].   
\ee
The free data is contained in the $\mathcal{O}(r^0)$ term in this expansion, which we may isolate using the saddle point approximation: 
\be
A_z^{(0)}(u,z,\bz)=-\frac{ie}{2(2\pi)^2}\p_z \hat{x}^i  \sum_{\alpha}\int_0^\infty d\omega_q[\epsilon_i^{*\alpha}a_\alpha(\omega_q \hat{x})e^{-i\omega_q u} - \epsilon_i^\alpha a_{\alpha}(\omega_q \hat{x})^\dagger e^{i\omega_q u}].   
\ee
To extract the contribution from the zero modes, we define the following operator:
\begin{align}\notag
	F^\omega_{u z}(z,\bz) &\equiv \int_{-\infty}^\infty du\  e^{i \omega u} \partial_u A^{(0)}_z(u,z,\bz)\\
 				& \notag=  -\frac{  e }{4 \pi}      
					 \partial_z \hat{x}^i \sum_{\alpha}   \int_0^\infty d \omega_q \  \omega_q
				\left[  \varepsilon^{* \alpha}_i a_\alpha(\omega_q \hat{x})   \delta(\omega - \omega_q) 
				+  \varepsilon^{ \alpha}_i a_\alpha(\omega_q \hat{x})^\dagger   \delta(\omega + \omega_q)\right] .  
\end{align}
We can separate this operator into its positive and negative frequency components
\be
F^\omega_{u z}(z,\bz)=- \frac{e \omega  }{4 \pi}  \partial_z \hat{x}^i      
				 \sum_{\alpha}\varepsilon^{* \alpha}_i a_\alpha(\omega  \hat{x})  , \;\;\;\;\; F^{-\omega}_{u z}(z,\bz)=- \frac{e \omega }{4 \pi}  \partial_z \hat{x}^i  \sum_{\alpha}    
				   \varepsilon^{ \alpha}_i a_\alpha(\omega \hat{x})^\dagger ,  
\ee
with $\omega > 0$ in both expressions. The zero mode is then given by
\begin{align}
			F^0_{u z}(z,\bz) &\equiv \tfrac{1}{2}\lim_{\omega \rightarrow 0} \left(F^\omega_{u z} + F^{-\omega}_{u z}\right)
				 = - \frac{e}{8 \pi}  \partial_z \hat{x}^i      \lim_{\omega\rightarrow 0}  \sum_{\alpha}\left[ 
				\omega \varepsilon^{* \alpha}_i a_\alpha(\omega  \hat{x})  
				+   \omega\varepsilon^{ \alpha}_i a_\alpha(\omega \hat{x})^\dagger  \right],  \label{eq:zeromode}
		\end{align}
and creates/annihilates soft photons. An analogous construction holds at $\mathcal{I}^-$ with the incoming soft photon operator given by
\begin{align}
	 F^0_{v z}(z,\bz)   
				& = \frac{e}{8 \pi}    \partial_z \hat{x}^i     \lim_{\omega\rightarrow 0} \sum_\alpha   \Big[ \omega 
				 \varepsilon^{* \alpha}_i a_\alpha(-\omega  \hat{x})  
				+      \omega \varepsilon^{ \alpha}_i a_\alpha(-\omega \hat{x})^\dagger\Big]    .  
 \end{align}

\subsection{Lienard-Wiechert Fields  }
	In the analysis that follows, we will need expressions for the electric field due to moving point charges, commonly known as Lienard-Wiechert fields.   
	The radial electric field due to a single particle of charge $eQ$, 
	moving with constant velocity $\vec{\beta}$ and  passing through the origin at $t = 0$ is given by
	\be
		\label{LWsingle}
		E_r(t,r, z, \bar{z}) =  \frac{Q e^2}{4 \pi } \frac{\gamma (r -t \hat{x}(z, \bz) \cdot \vec{\beta})}{  |\gamma^2[t- r \hat{x}(z, \bz) \cdot \vec{\beta}]^2 - t^2 + r^2|^{3/2}}
		.
	\ee
	Here $\hat{x}(z, \bar{z})$ is a unit vector specifying a point on the sphere and $\gamma^{-2}=1-\beta^2$. 
	
	The Lienard-Wiechert field near $\mathcal{I}^+$ due to a set of particles, each with charge $eQ_k$ and  moving with constant velocity $\vec{\beta}_k$ is derived by taking a superposition of the 
	single-particle fields (\ref{LWsingle}), writing them in retarded coordinates, and taking the large-$r$ limit with $u=t-r$ held fixed
	\be \label{eplus}
		E^{+}_r(z, \bar{z})=\sum_k\frac{Q_ke^2}{4 \pi \gamma_k^2 r^2}\frac{1}{[1-\hat{x}(z,\bar{z})\cdot \vec{\beta_k}]^2}. 
	\ee
Likewise, the field near $\mathcal{I}^-$ is derived by taking the large-$r$ limit of the field (\ref{LWsingle}) in advanced coordinates with fixed $v$
	\be\label{daz}
		E_r^{-}(z, \bar{z})=\sum_k\frac{Q_ke^2}{4 \pi \gamma_k^2 r^2}\frac{1}{[1+\hat{x}(z, \bz)\cdot \vec{\beta_k}]^2}.
	\ee

Importantly,  the Lienard-Wiechert formula (\ref{LWsingle}) implies that the value of $E_r$ near spatial infinity $i^0$ depends on how it is approached.  In particular, $E^+_r$ and $E^-_r$ at a fixed angle from the origin are not in general equal near $i^0$: rather they obey the antipodal matching condition  (\ref{matchF}).\footnote{The  constant-velocity trajectory considered in this section 
	gives rise to a Lienard-Wiechert field that is
	insensitive to the choice of Green's function.  In the appendix, we consider slightly more complicated trajectories to demonstrate that this matching condition holds for a generic 
	Green's function.}

	For unaccelerated charges, the asymptotic electric field and the asymptotic magnetic field $\vec{B} = \hat{x} \times \vec{E}$ are time-independent. 
Since the ``hard" radiative photons involved in the scattering process exit/enter $\mathcal{I}^\pm$ at finite values of retarded/advanced time, the electromagnetic fields at $\mathcal{I}^+_+$ and $\mathcal{I}^-_-$  arise solely from the collection of  charged particles long after/before the scattering process occurs and thus are of the form given above.

\section{Symmetries of the $\mathcal{S}$-matrix}
In this section we determine the phase associated to a large gauge transformation on a (dressed) asymptotic massive charged particle state, find the 
$\mathcal{S}$-matrix Ward identity and finally demonstrate its equivalence to the soft photon theorem. 
\subsection{Gauge transformations of asymptotic states  }

Outgoing massless particles of charge $eQ$ and momentum $p$, as considered in \cite{He:2014cra}, pierce $\ci^+$ at a definite point $(z(p),\bz(p))$.  The associated out-state therefore acquires a phase 
\be \label{mnv} |p \rangle_{out} \to e^{iQ\e(z(p),\bz(p))}| p \rangle_{out}\ee under a large gauge transformation.
Here we are interested in massive particles that never reach $\ci^+$, so determining the associated phase is more subtle. There is no canonical  point on the $S^2$ associated to a massive particle\footnote{Since the wavefunction of a massless particle localizes to a point on the  conformal sphere at null infinity, it is related to a local operator insertion on that sphere.  Massive particles, on the other hand, will involve non-local weighted  integrals of operators over the sphere. }. Indeed 
 a massive particle with zero three-momentum is rotationally invariant. In this subsection, we use the Lienard-Wiechert formula to
 determine the analog of the phase (\ref{mnv}).
 The asymptotic states associated to the QED $\mathcal{S}$-matrix are typically taken to be free photons and ``bare" non-interacting charged particles.  However, infrared (IR) divergences entering 1-loop calculations render such matrix elements formally infinite. Unlike ultraviolet divergences, which are associated to high energy, unobservable physics, these IR divergences have a clear physical interpretation associated to low energy observables.   
This well-understood problem stems from an improper choice of asymptotic scattering states\cite{Dirac:1955uv,Bloch:1937pw, Kulish:1970ut,Chung:1965zza, Zwanziger:1973if , Kibble:1969ip, Kibble:1969ep,Kibble:1969kd}. Formally, the standard assumption that the asymptotic dynamics is governed by the free Hamiltonian $H_0$ does not hold in QED. It fails precisely because the long-wavelength, infrared sector of the gauge theory fails to decouple from the charged particles at early/late times.  Physically, it is impossible to separate a charged particle from its electromagnetic field. The ability to do so would violate Gauss' law, which can be implemented as a constraint in the quantum theory:
    \be \label{gauss}
\left[ \nabla \cdot \vec{E}(x,t) -e^2 \rho(x,t) \right] |phys \rangle =0 .
\ee
The resolution, worked out in \cite{Dirac:1955uv,Bloch:1937pw, Kulish:1970ut,Chung:1965zza, Zwanziger:1973if , Kibble:1969ip, Kibble:1969ep,Kibble:1969kd}, is to choose scattering states that diagonalize the asymptotic Hamiltonian $H_{as}$, defined to account for the long-range interactions in QED. 
In particular, ``undressed" charged particles without electric fields are not eigenstates of the asymptotic Hamiltonian and cannot appear as scattering states. Consider the  case of a single ``undressed" electron state 
\be
\psi(x)|0\rangle.
\ee
This state is not invariant under small gauge transformations and does not satisfy the Gauss constraint. To form a physical state, we must also include the electron's electric field.  This problem can be solved by introducing dressed states of the general form \cite{Dirac:1955uv,Bagan:1999jf,Bagan:2001uq,Giddings:2015lla}
\be \label{physicalstate}
|\psi(x),E(x)\rangle=\psi(x)\exp\left[{\frac{i}{e^2}\int E_i(x,y)A^i(y)d^3y}\right]|0\rangle
\ee
  for any $E_i(x,y)$ satisfying the Gauss constraint
  \be \label{constraint}
  \nabla^iE_i(x,y)=Qe^2\delta^3(x-y).
  \ee
Under a general gauge transformation, this state transforms as
  \be
   |\psi(x),E(x)\rangle\to \exp \left[ \frac{i}{e^2}\int_{S^2_{\infty}}d^2z {\gamma}_{z\bz} r^2 \epsilon(z, \bz)   E_r (z, \bz)  \right]  |\psi(x),E(x)\rangle. 
  \ee
  For small gauge transformations $\epsilon(x)$ vanishing at infinity, the state is left invariant, while large gauge transformations rotate the state by a phase. 
  
Suppose  that the outgoing states with momentum $p^\mu= \gamma m [ 1, \vec{\beta} ]$
have been chosen to diagonalize the asymptotic Hamiltonian and are dressed with their Lienard-Wiechert fields so that the electric field at $\mathcal{I}^+_+$ is given by 
\be \label{fks}
E_r(r,z, \bar{z})|p\rangle_{out}=\left[ \frac{Qe^2}{4 \pi \gamma^2 r^2}\frac{1}{[1-\hat{x}(z,\bar{z})\cdot \vec{\beta}]^2}\right] |p\rangle_{out}.\ee
Under a large gauge transformation, such a state acquires a phase 
\be \label{phase}
|p\rangle_{out} \to \exp \left[ i\int_{\mathcal{I}^+_+}d^2z {\gamma}_{z\bz} \epsilon^+\left( \frac{1}{4 \pi \gamma^2 }\frac{Q}{[1-\hat{x}(z,\bar{z})\cdot \vec{\beta}]^2}\right) \right]|p\rangle_{out}. \ee
Similarly, in-states transform as
\be
|p\rangle_{in} \to \exp \left[ i\int_{\mathcal{I}^-_-}d^2z {\gamma}_{z\bz} \epsilon^-\left( \frac{1}{4 \pi \gamma^2 }\frac{Q}{[1+\hat{x}(z,\bar{z})\cdot \vec{\beta}]^2}\right) \right]|p\rangle_{in}. 
\ee 
For an $n$-particle state, the phase will be a sum of $n$ such terms. This phase replaces the much simpler expression (\ref{mnv}) for massless particles but nevertheless, as will be seen shortly,   precisely reproduces the soft factor for massive particles.  

In principle, a state could be dressed with a different solution to the Gauss constraint.  However, any such solution can only differ from the Lienard-Wiechert solution by a solution to the source-free  Maxwell equations.  
These more general radiative solutions are eliminated by our boundary conditions at $\mathcal{I}^+_+$ and $\mathcal{I}^-_-$, which single out the sourced, non-radiative contribution to the field.  Physically, we want to dress our states with a `minimal' photon cloud that satisfies the Gauss constraint but excludes additional photons corresponding to radiative contributions to the electromagnetic field.   These radiative contributions cross $\mathcal{I}$ and do not pass through $i^{\pm}$.

\subsection{Ward Identity  } \label{section32}
We are now in a position to discuss the symmetries of the $\mathcal{S}$-matrix. The symmetry transformations (\ref{gaugetransform})  for massless matter fields have already been analyzed in  \cite{He:2014cra}, where it was demonstrated that the charge
\be
	Q_\varepsilon^+ = \frac{1}{e^2} \int_{ \mathcal{I}_-^+} d^2z  \gamma_{z\bz}   \varepsilon^+(z,\bz)    F^{(2)}_{ru}(z,\bz)   
\ee  
generates the correct $\mathcal{I}^+$ symmetry transformation on the gauge field and matter fields. The form of this charge is essentially fixed by the transformation law for the gauge field.
We can use the leading order Maxwell equation (\ref{evolution}) to turn this expression into an integral over $\mathcal{I}^+$.  As discussed in section 2.1,  the existence of massive particles generates charge flux through future timelike infinity, so the local charge operator takes the form  
\begin{align} \label{qplus}
	Q_\varepsilon^+   \notag 
		&= \frac{1}{e^2} \int_{\mathcal{I}^+} \gamma_{z\bz}  du d^2z\    \epsilon^+   \partial_u (D^zA^{(0)}_z+D^\bz A^{(0)}_\bz)+ \frac{1}{e^2} \int_{\mathcal{I}_+^+}d^2z \  \gamma_{z\bz}  \varepsilon^+   F^{(2)}_{ru}\\
		&\equiv\frac{1}{e^2} \int_{S^2} \gamma_{z\bz}   d^2z\     \epsilon^+   (D^zF^0_{uz}+D^\bz F^0_{u\bz}) +\frac{1}{e^2} \int_{\mathcal{I}_+^+}d^2z \  \gamma_{z\bz}   \epsilon^+    F^{(2)}_{ru}.  
\end{align}  

The first piece of the charge is written in terms of the soft photon operator and will be referred to as the soft charge $Q_S^+$.  If we consider the fixed-angle charge by choosing 
$\epsilon(z, \bar{z}) = \delta^2(z - w)$, then the second term is simply the radial electric field in the direction $(w, \bar{w})$ resulting from the charged particles in the final state at $i^+$. We label this term the hard charge $Q_H^+$. It differs from the expression for the hard charge in the massless case which involves an integral over $\ci^+$. 

An analogous computation can be performed at $\mathcal{I}^-$, where the hard charge encodes the radial electric field of the charged particles in the initial state at $i^-$. The charge is given by
\begin{align} \label{qminus}
	Q_\varepsilon^-    
		&=\frac{1}{e^2} \int_{S^2} \gamma_{z\bz}   d^2z\    \epsilon^-  (D^zF^0_{vz}+D^\bz F^0_{v\bz})+ \frac{1}{e^2} \int_{\mathcal{I}_-^-}d^2z \  \gamma_{z\bz}  \varepsilon^-   F^{(2)}_{rv} \equiv Q_S^-+Q_H^-.
\end{align}  

The statement that the transformations (\ref{gaugetransform}) are symmetries of the $\mathcal{S}$-matrix is equivalent to the statement that the charges (\ref{qplus}) and (\ref{qminus})  commute with the $\mathcal{S}$-matrix:
\begin{align}
	 \<\text{out}|  \left(Q_\varepsilon^+ \mathcal{S} - \mathcal{S} Q^-_\varepsilon\right) | \text{in} \> = 0.
\end{align}
In order to facilitate comparison with the soft theorem, we separate the hard and soft contributions and rearrange the Ward identity:
\begin{align} \label{Ward}
	 \<\text{out}|  \left(Q_S^+ \mathcal{S} - \mathcal{S} Q^-_S\right) | \text{in} \>  =  - \<\text{out}|  \left(Q_H^+ \mathcal{S} - \mathcal{S} Q^-_H\right) | \text{in} \>.
\end{align}

\subsection{Soft Theorem $\to$ Ward Identity}

The soft photon theorem for the emission of an outgoing photon in a scattering process with $m$ incoming hard particles and $(n-m)$ outgoing hard particles reads
\be
\lim_{\omega \to 0} \omega \langle p_{m+1},\dots  | a_{\alpha}(q) \mathcal{S}| p_1, \dots \rangle
	 =e\omega \left[ \sum_{k=m+1}^{n} Q_k \frac{p_k \cdot \epsilon_{\alpha}}{p_k \cdot q} - \sum_{k=1}^{m} Q_k\frac{p_k\cdot \epsilon_\alpha}{p_k \cdot q}   \right]\langle p_{m+1}, \dots  |  \mathcal{S}| p_1, \dots \rangle.  
	 \label{eq:softphoton}
\ee
A null momentum vector is uniquely specified by an energy and a point $z$ on the asymptotic sphere, and so we parameterize the photon's momentum as
\be
q^\mu=\omega[1,\hat{x}(z,\bz)]\equiv \omega \hat{q}^\mu(z,\bz),  
\ee
where $\hat{x}: S^2 \to \mathbb{R}^3$ is an embedding of the sphere into flat three-dimensional space. 

We parameterize a massive particle's momentum as
\be
p_k^\mu= \gamma_k m_k [ 1, \vec{\beta}_k],
\ee
where $m$ is the rest mass of the particle, $\vec{\beta}$ is  the particle's velocity and $\gamma$ is the relativistic factor $\gamma^{-2} = 1- \beta^2 $.

We can relate the left-hand side of equation (\ref{eq:softphoton}) to the zero mode operator defined in equation (\ref{eq:zeromode}) by taking a weighted sum over polarizations.  If we perform the analogous operation on the 
right-hand side and use the identity
\begin{align}
			\partial_z \hat{x}^i(z,\bz)   \sum_\alpha \varepsilon^{*\alpha}_i   \frac{\ p_k \cdot \varepsilon_\alpha}{p_k \cdot \hat{q}(z, \bz)}
				&=\partial_z \log(p_k \cdot \hat{q}),  
 \end{align} 
 the soft theorem can be written
 \begin{align}
			  \<p_{m+1},\dots| F^0_{u z}  \mathcal{S} |p_1,\dots\>
				 = -\frac{e^2}{8 \pi}   \left[\sum_{k=m+1}^n   Q_k\ \partial_z \log(p_k \cdot \hat{q})
					-\sum_{k=1}^m  Q_k\ \partial_z \log(p_k \cdot \hat{q})\right]\<p_{m+1},\dots|  \mathcal{S} |p_1,\dots\> .  
		\end{align}
		
	The soft photon theorem for an incoming soft photon reads
\be
\lim_{\omega \to 0} \omega \langle p_{m+1}, \dots  |  \mathcal{S} a_{\alpha}(q)^\dagger| p_1, \dots \rangle
	 =-e \omega\left[ \sum_{k=m+1}^{n} Q_k \frac{p_k \cdot \epsilon^*_{\alpha}}{p_k \cdot q} - \sum_{k=1}^{m} Q_k\frac{p_k\cdot \epsilon^*_\alpha}{p_k \cdot q}   \right]\langle p_{m+1}, \dots  |  \mathcal{S}| p_1, \dots \rangle.  
	 \label{eq:softphotonIN}
\ee 
An identical calculation yields		
	 \begin{align}
			  \<p_{m+1},\dots|  \mathcal{S}F^0_{v z}  |p_1,\dots \>
				 = \frac{e^2}{8 \pi}   \left[\sum_{k=m+1}^n   Q_k\ \partial_z \log(p_k \cdot \hat{q}')
					-\sum_{k=1}^m  Q_k\ \partial_z \log(p_k \cdot \hat{q}')\right]\<p_{m+1},\dots|  \mathcal{S} |p_1,\dots\>,   
		\end{align}
where  $\hat{q}' = [1,-\hat{x}^i(z,\bz)]$.	
Taking the divergence of each equation, using global charge conservation, and integrating against the respective gauge parameter, we find

	 \begin{align}
			  \<p_{m+1},\dots|&  \left(\int_{S^2} d^2 z \gamma_{z \bz} \varepsilon^+ ( D^zF^0_{u z}+ D^\bz F^0_{u \bz})\right) \mathcal{S} |p_1,\dots\> \notag \\&
				 =  - \frac{1 }{2 } \int_{S^2} d^2z \gamma_{z \bz}   \varepsilon^+  r^2\Big( [E^{+ }_r]_{\text{out}}-[E^{+ }_r]_{\text{in}}\Big) \<p_{m+1},\dots|  \mathcal{S} |p_1,\dots\>  
	 \end{align}
	 and
	  \begin{align}
			  \<p_{m+1},\dots|& \mathcal{S} \left(\int_{S^2} d^2 z \gamma_{z \bz} \varepsilon^- ( D^zF^0_{vz}+ D^\bz F^0_{v \bz})\right)  |p_1,\dots\> \notag \\&
				 =   \frac{1 }{2 } \int_{S^2} d^2z \gamma_{z \bz}   \varepsilon^-  r^2\Big( [E^{- }_r]_{\text{out}}-[E^{- }_r]_{\text{in}}\Big) \<p_{m+1},\dots|  \mathcal{S} |p_1,\dots\>.  
	 \end{align} Taking the difference and using the matching conditions (\ref{matchF})-(\ref{matchEp}), we obtain
	 \begin{align}
	 	\<p_{m+1},\dots|& Q^+_S \mathcal{S}  - \mathcal{S} Q^-_S |p_1,\dots\> \notag  \\&
				 =  -  \frac{1}{e^2}  \int_{S^2} d^2z  \gamma_{z \bz} r^2\Big(  \varepsilon^+ E_r \big|_{\mathcal{I}^+_+} -\varepsilon^- E_r \big|_{\mathcal{I}^-_-} \Big)\<p_{m+1},\dots|\mathcal{S} |p_1,\dots\> \\
		&=	-\<p_{m+1},\dots|Q_H^+\mathcal{S}-\mathcal{S} Q_H^- |p_1,\dots\>	 .
	 \end{align}	
	 This precisely reproduces the Ward identity  (\ref{Ward}).
	 
	 In conclusion, while the details are more intricate than the massless case, the soft photon theorem is the Ward identity of an infinite-dimensional asymptotic symmetry group for abelian gauge theories with massive particles. We expect similar conclusions apply to other contexts such as non-abelian gauge theory and gravity.   

\section*{Acknowledgements}
We are  grateful  to D. Christodoulou, D. Farhi, I. Feige, D. Garfinkle, T. He, V. Lysov, P. Mitra, S. Pasterski,  M. Schwartz and A. Zhiboedov for useful conversations. This work was supported in part by DOE grant DE-FG02-91ER40654 and the Fundamental Laws Initiative at Harvard.

\section{Appendix: Gauge field strength near $i^0$}
	
	In this section, we consider an idealized semiclassical scattering process in which $m$ incoming massive particles with constant velocities $\{ \vec{\beta}_1, \dots ,\vec{\beta}_m\}$ scatter to $(n-m)$ outgoing massive particles with constant velocities $\{ \vec{\beta}_{m+1},\dots, \vec{\beta}_{n} \}$.  For scattering occurring at the origin at $t = 0$, the semiclassical electromagnetic current is given by
	\begin{align}
		j_\mu(x)&=\sum_{k=1}^m Q_k\int d\tau   U^k_{\mu}\Theta(-\tau)\delta^{(4)}(x-U^k\tau) + \sum_{k=m+1}^n Q_k\int  d\tau   U^k_{\mu}\Theta(\tau) \delta^{(4)}(x-U^k\tau), 
\end{align}
where $U^k_\mu = \gamma_k[1, \vec{\beta}_k]$ is the 4-velocity of the $k^{th}$ particle.  
Ignoring  the radiative contributions arising from the infinite acceleration of particles at the origin, the field strength sourced by this current takes the form
\begin{align}
	F_{rt}(x) = \frac{ e^2}{4 \pi }\sum_{k=1}^m  \frac{g(x;\vec{\beta}_k)\ Q_k\gamma_k (r -t \hat{x}  \cdot \vec{\beta}_k)}{  |\gamma_k^2[t- r \hat{x}  \cdot \vec{\beta}_k]^2 - t^2 + r^2|^{3/2}}
			+ \frac{ e^2}{4 \pi } \sum_{k=m+1}^n\frac{h(x; \vec{\beta}_k)\ Q_k\gamma_k (r -t \hat{x}  \cdot \vec{\beta}_k)}{  |\gamma_k^2[t- r \hat{x}  \cdot \vec{\beta}_k]^2 - t^2 + r^2|^{3/2}},
\end{align}
where the functional form of $g$ and $h$ depends on the choice of Green's function.  

For the retarded solution, the asymptotic behavior of $g$ and $h$  is given by
\begin{align}	
	&g(r = \infty, u, \hat{x};\vec{\beta}_k) = \Theta(-u), && 
	h(r = \infty,u, \hat{x};\vec{\beta}_k) = \Theta(u), \\
	&g(r = \infty,v, \hat{x};\vec{\beta}_k) = 1,&& 
	h(r = \infty,v,\hat{x};\vec{\beta}_k) = 0.
\end{align}
The electric field at $\mathcal{I}^+_-$ is obtained by working in retarded coordinates and taking the limit $r \rightarrow \infty$, followed by the limit $u \rightarrow -\infty$.  This electric field will be of the form (\ref{eplus}), but only receives contributions from the incoming particles.  On the other hand, the electric field at $\mathcal{I}^-_+$ is obtained by working in advanced coordinates and taking the limit $r \rightarrow \infty$, followed by the limit $v \rightarrow +\infty$.  The electric field measured at $\mathcal{I}^-_+$ will be of the form (\ref{daz}), but will also only receive contributions from the incoming particles, thereby satisfying the matching condition (\ref{matchF}).

Likewise, for the advanced solution, the asymptotic behavior of $g$ and $h$  is given by
\begin{align}	
	&g(r = \infty, u,   \hat{x};\vec{\beta}_k) = 0, && 
	h(r = \infty,u,   \hat{x};\vec{\beta}_k) = 1, \\
	&g(r = \infty,v,  \hat{x};\vec{\beta}_k) = \Theta(-v),&& 
	h(r = \infty,v,  \hat{x};\vec{\beta}_k) = \Theta(v).
\end{align}
Hence, the advanced solution also obeys the matching condition (\ref{matchF}) near $i^0$, but in contrast to the retarded solution, only receives contributions from the outgoing particles.  Moreover,
linear combinations of the advanced and retarded solutions evaluated near $i^0$ will obey the matching condition and receive contributions from both outgoing and incoming particles.

Of course, one could always add a homogeneous solution to the free Maxwell equations which does not obey the matching condition. However,  we do not know of any physical application in which it is natural to do so: finite energy wave packets die off at $i^0$.  Hence, we conclude that 
the antipodal matching condition (\ref{matchF}) holds in generic physical applications.

\end{document}